\documentclass[12pt]{article}

\usepackage{amssymb,amsmath,latexsym,amsthm,amsfonts}
\usepackage{graphicx}
\usepackage{mathrsfs}
\usepackage{natbib}
\bibpunct{(}{)}{,}{a}{,}{,}

\usepackage{geometry}
\usepackage{setspace} 

\newcommand{\ds}{\displaystyle}
\newcommand{\ev}{m}

\newcommand{\bx}{\mathbf{x}}
\newcommand{\bX}{\mathbf{X}}
\newcommand{\by}{\mathbf{y}}
\newcommand{\bz}{\mathbf{z}}
\newcommand{\btheta}{{\boldsymbol \theta}}

\newenvironment{theacknowledgments}
     {\section*{Acknowledgements}}
     {\par}

\bibpunct{(}{)}{,}{a}{}{,}

\title{{\bf Importance sampling methods for Bayesian discrimination between embedded models}}

\author{
{{\sc J.-M.~Marin}$^{1,3,}$\thanks{Institut de Math\'ematiques et Mod\'elisation de Montpellier,
Universit\'e Montpellier 2, Case Courrier 51, 34095 Montpellier cedex 5, France. 
Email: \texttt{Jean-Michel.Marin@univ-montp2.fr}}}
and
{{\sc C.P.~Robert}$^{2,3,}$\thanks{CEREMADE - Universit\'e Paris Dauphine, 75775 Paris, and CREST, INSEE,
Paris, France.  Email: \texttt{xian@ceremade.dauphine.fr}}}\\
$^1$Institut de Math\'ematiques et Mod\'elisation de Montpellier,\\
$^2$Universit\'e Paris Dauphine, and\\ $^3$Centre de Recherche en Economie et Statistique, INSEE, Paris
}

\begin{document}

\maketitle
% \doublespacing

\begin{abstract}
This paper surveys some well-established approaches on the approximation of Bayes factors
used in Bayesian model choice, mostly as covered in \cite{chen:shao:ibrahim:2000}. Our focus here is
on methods that are based on importance sampling strategies---rather than variable dimension
techniques like reversible jump MCMC---, including:
crude Monte Carlo, maximum likelihood based importance sampling, bridge and harmonic mean sampling, as well as 
Chib's method based on the exploitation of a functional equality. We demonstrate in this 
survey how these different methods can be efficiently implemented for testing the significance of a 
predictive variable in a probit model. Finally, we compare their performances on a real dataset.

\end{abstract}

\section{Introduction}

The contribution of Jim Berger to the better understanding of Bayesian testing is fundamental and wide-ranging,
from establishing the fundamental difficulties with $p$-values in \cite{berger:sellke:1987} to formalising
the intrinsic Bayes factors in \cite{berger:pericchi:1996}, to solving the difficulty with improper priors 
in \cite{berger:pericchi:varshavsky:1998}, and beyond! While our contribution in this area is obviously much more limited,
we aim at presenting here the most standard approaches to the approximation of Bayes factors.

The Bayes factor indeed is a fundamental procedure that stands at the core of the Bayesian theory of testing 
hypotheses, at least in the approach advocated by both \cite{jeffreys:1939} and by \cite{jaynes:2003}.
(Note that \citealp{robert:chopin:rousseau:2009}, provides a reassessment of the crucial role of 
\citealp{jeffreys:1939} in setting a formal framework for Bayesian testing as well as for regular inference.)
Given an hypothesis $H_0:\,\theta\in\Theta_0$ on the parameter $\theta\in\Theta$ of a statistical model,
with observation $y$ and density $f(y|\theta)$, under a compatible prior of the form
$$
\pi(\Theta_0) \pi_0(\theta) + \pi(\Theta_0^c) \pi_1(\theta)\,,
$$
the {\em Bayes factor} is defined as the posterior odds to prior odds ratio, namely
$$%\begin{eqnarray*}
B_{01}(y)  =  \displaystyle{\frac{\pi(\Theta_0|y)}{ \pi(\Theta_0^c|y)}\bigg/\frac{\pi(\Theta_0)}
{\pi(\Theta_0^c)}} =  {\displaystyle{\int_{\Theta_0} f(y|\theta) \pi_0(\theta) \text{d}\theta}}\bigg/ {
       \displaystyle{\int_{\Theta_0^c} f(y|\theta) \pi_1(\theta) \text{d}\theta} }\,.
$$%\end{eqnarray*}
Model choice can be considered from a similar perspective, since, under the Bayesian paradigm (see,
e.g., \citealp{robert:2001}), the comparison of models 
$$
\mathfrak{M}_i : y \sim f_i(y|\theta_i), \quad\theta_i \sim \pi_i(\theta_i), \quad\theta_i\in\Theta_i , \quad i \in \mathfrak{I}\,,
$$
where the family $\mathfrak{I}$ can be finite or infinite, leads to posterior probabilities 
of the models under comparison such that
$$
\mathbb{P}\left(\mathfrak{M}=\mathfrak{M}_i|y\right)\propto p_i \int_{\Theta_i} 
f_i(y|\theta_i)\pi_i(\theta_i) \text{d}\theta_i\,,
$$
where $p_i=\mathbb{P}(\mathfrak{M}=\mathfrak{M}_i)$ is the prior probability of model $\mathfrak{M}_i$.

In this short survey, we consider some of the most common Monte Carlo solutions used to approximate a generic
Bayes factor or its fundamental component, the {\em evidence}
$$ 
\ev_i = \int_{\Theta_i} \pi_i(\theta_i) f_i(y|\theta_i)\,\text{d}\theta_i\,,
$$
aka the marginal likelihood. Longer entries can be found in \cite{carlin:chib:1995},
\cite{chen:shao:ibrahim:2000}, \cite{robert:casella:2004}, or \cite{friel:pettitt:2008}.
Note that we only briefly mention here trans-dimensional methods issued from the revolutionary paper
of \cite{green:1995}, since our goal is to demonstrate that within-model simulation methods allow for 
the computation of Bayes factors and thus avoids the additional complexity involved in trans-dimensional 
methods. While ameanable to an importance sampling technique of sorts,
the alternative approach of nested sampling \citep{skilling:2007a} is discussed in 
\cite{chopin:robert:2007} and \cite{robert:wraith:2009}.

\section{The Pima Indian benchmark model}
In order to compare the performances of all methods presented in this survey, 
we chose to evaluate the corresponding estimates of the Bayes factor
in the setting of a single variable selection for a probit model and to repeat
the estimation in a Monte Carlo experiment to empirically assess the variability
of those estimates. 

We recall that a probit model can be represented as a natural latent variable model
in that, if we consider a sample $z_1,\ldots,z_n$ of $n$ independent latent variables 
associated with a standard regression model, i.e.~such that
$z_i|\btheta\sim\mathcal{N}\left(\bx_i^\text{T}\btheta,1\right)$, 
where the $\bx_i$'s are $p$-dimensional covariates and $\btheta$ is the vector of regression coefficients,
%$\bx_i^\text{T}$ is the transpose of $\bx_i$ and $\btheta$ is an unknown vector of $\mathbb{R}^p$) and 
then $y_1,\ldots,y_n$ such that
% the observed binary responses such that
$$
%y_i=\left\{\begin{array}{ll}
 %%1 & \mbox{if } z_i>0 \\
 %%0 & \mbox{otherwise}
%\end{array}\right..
y_i = \mathbb{I}_{z_i>0}
$$
is a probit sample. Indeed, given $\btheta$, the $y_i$'s are independent Bernoulli rv's with
$\mathbb{P}(y_i=1|\btheta)=\Phi\left(\bx_i^\text{T}\btheta\right)$ where $\Phi$
is the standard normal cdf.
%cumulative distribution function. \\

The choice of a reference prior distribution for the probit model is open to debate,
but the connection with the latent regression model induced \cite{marin:robert:2007}
to suggest a $g$-prior model, $\btheta\sim\mathcal{N}\left(0_p,n(\bX^\text{T}\bX)^{-1}\right)$,
with $n$ as the $g$ factor and $\bX$ as the regressor matrix. 
The corresponding posterior distribution is then associated
%as prior distribution on $\btheta$ ($\bX=\left[\bx_1\left|\bx_2\left|\ldots\left|\bx_p\right.\right.\right.\right]$),
with the density 
%of the posterior distribution is such that
\begin{equation}
\pi(\btheta|\by,\bX)\propto \prod_{i=1}^n 
\left\{1-\Phi\left(\bx_i^\text{T}\btheta\right)\right\}^{1-y_i}
\Phi\left(\bx_i^\text{T}\btheta\right)^{y_i} 
\times
\exp\left\{-\btheta^\text{T}(\bX^\text{T}\bX)\btheta/2n\right\}\,,
\label{eq:postprobit}
\end{equation}
where $\by=(y_1,\ldots,y_n)$. In the completed model, i.e.~when including the latent variables 
$\bz=\left(z_1,\ldots,z_n\right)$ into
the model, the $y_i$'s are deterministic functions of the $z_i$'s and the so-called completed likelihood
is
$$
f(\by,\bz|\btheta)=(2\pi)^{-n/2}\exp\left(-\sum_{i=1}^n\left(z_i-\bx_i^\text{T}\btheta\right)^2/2\right)
\prod_{i=1}^n\left(\mathbb{I}_{y_i=0}\mathbb{I}_{z_i\leq 0}+\mathbb{I}_{y_i=1}\mathbb{I}_{z_i>0}\right)\,.
$$
The derived conditional distributions
\begin{equation}
z_i|y_i,\theta\sim\left\{\begin{array}{ll}
\mathcal{N}_+\left(\bx_i^\text{T}\btheta,1,0\right) & \mbox{if}\quad y_i=1\,, \\
\mathcal{N}_-\left(\bx_i^\text{T}\btheta,1,0\right) & \mbox{if}\quad y_i=0\,,
\end{array}\right.
\label{gibbs1}
\end{equation}
are of interest for constructing a Gibbs sampler on the completed model,
where $\mathcal{N}_+\left(\bx_i^\text{T}\btheta,\allowbreak 1,\allowbreak 0\right)$ denotes the Gaussian
distribution with mean $\bx_i^\text{T}\btheta$ and variance $1$ that is left-truncated at $0$, while
$\mathcal{N}_-\left(x_i^\text{T}\theta,1,0\right)$ denotes the 
symmetrical normal distribution that is right-truncated at $0$. The corresponding full conditional on the parameters
is given by
\begin{equation}
\btheta|\by,\bz\sim\mathcal{N}\left(\frac{n}{n+1}(\bX^\text{T}\bX)^{-1}\bX^\text{T}\bz,\frac{n}{n+1}(\bX^\text{T}\bX)^{-1}\right)\,.
\label{gibbs2}
\end{equation}
Indeed, since direct simulation from
the posterior distribution of $\btheta$ is intractable, \cite{albert:chib:1993b} suggest implementing 
a Gibbs sampler based on the above set of full conditionals. More precisely,
given the current value of $\btheta$, one cycle of the Gibbs algorithm 
produces a new value for $\bz$  as simulated from the
conditional distribution (\ref{gibbs1}), which, 
when substituted into (\ref{gibbs2}), produces a new value for $\btheta$.
Although it does not impact the long-term properties of the sampler,
the starting value of $\btheta$ may be taken as the maximum likelihood estimate
to avoid burning steps in the Gibbs sampler. 

Given this probit model, the dataset we consider covers a population of women who were at least 21 years old,
of Pima Indian heritage and living near Phoenix, Arizona. These women were tested for diabetes 
according to World Health Organization (WHO) criteria. The data were collected by the US National 
Institute of Diabetes and Digestive and Kidney Diseases, and is available with the basic {\sf R} package
\citep{rmanual}. This dataset, used as a benchmark for supervised learning methods, contains information
about $332$ women with the following variables:
\begin{itemize}
\item[--] \verb+glu+: plasma glucose concentration in an oral glucose tolerance test;
\item[--] \verb+bp+: diastolic blood pressure (mm Hg);
\item[--] \verb+ped+: diabetes pedigree function;
\item[--] \verb+type+: Yes or No, for diabetic according to WHO criteria.
\end{itemize} 
For this dataset, the goal is to explain the diabetes variable \verb+type+ by using the 
explanatory variables \verb+glu+, \verb+bp+ and \verb+ped+. The following table is an
illustration of a classical (maximum likelihood) analysis of this
dataset, obtained using the {\sf R} {\sf glm()} function with the probit link:
% \singlespacing
\begin{verbatim}
Deviance Residuals: 
    Min       1Q   Median       3Q      Max  
-2.1347  -0.9217  -0.6963   0.9959   2.3235  
Coefficients:
     Estimate Std. Error z value Pr(>|z|)    
glu  0.012616   0.002406   5.244 1.57e-07 ***
bp  -0.029050   0.004094  -7.096 1.28e-12 ***
ped  0.350301   0.208806   1.678   0.0934 .  
---
Signif. codes:  0 '***' 0.001 '**' 0.01 '*' 0.05 '.' 0.1 ' ' 1 
(Dispersion parameter for binomial family taken to be 1)

    Null deviance: 460.25  on 332  degrees of freedom
Residual deviance: 386.73  on 329  degrees of freedom
AIC: 392.73
Number of Fisher Scoring iterations: 4
\end{verbatim}
% \doublespacing
This analysis sheds some doubt on the relevance of the covariate \verb+ped+ in the model
and we can reproduce the study from a Bayesian perspective, computing the
Bayes factor $B_{01}$ opposing the probit model only based on
the covariates \verb+glu+ and \verb+bp+ (model 0) to the probit model based on the covariates
\verb+glu+, \verb+bp+, and \verb+ped+ (model 1). This is equivalent to testing the hypothesis
$H_0:\theta_3=0$ since the models are nested,
where $\theta_3$ is the parameter of the probit model associated with covariate \verb+ped+. 
(Note that there is no intercept in either model.) If we denote by
$\bX_0$ the $332\times 2$ matrix containing the values of \verb+glu+ and \verb+bp+
for the $332$ individuals and by $\bX_1$ the $332\times 3$ matrix containing the values of
the covariates \verb+glu+, \verb+bp+, and \verb+ped+, the Bayes factor $B_{01}$ is given by
\begin{align}\label{eq:bfprobit}
(2\pi)^{1/2}n^{1/2}&\frac{|(\bX_0^\text{T}\bX_0)|^{-1/2}}{|(\bX_1^\text{T}\bX_1)|^{-1/2}}\\
&\frac{\ds \int_{\mathbb{R}^2}\prod_{i=1}^n \{1-\Phi\left((\bX_0)_{i,\cdot}\btheta\right)\}^{1-y_i}
\Phi\left((\bX_0)_{i,\cdot}\btheta\right)^{y_i}
\exp\left\{-\btheta^\text{T}(\bX_0^\text{T}\bX_0)\btheta/2n\right\}\text{d}\btheta}
{\ds \int_{\mathbb{R}^3}\prod_{i=1}^n \{1-\Phi\left(\bX_1)_{i,\cdot}\btheta\right)\}^{1-y_i}
\Phi\left(\bX_1)_{i,\cdot}\btheta\right)^{y_i}
\exp\left\{-\btheta^\text{T}(\bX_1^\text{T}\bX_1)\btheta/2n\right\}\text{d}\btheta} \nonumber\\
&=\frac{\ds \mathbb{E}_{\mathcal{N}_2(0_2,n(\bX_0^\text{T}\bX_0)^{-1})}\left[\prod_{i=1}^n 
\{1-\Phi\left((\bX_0)_{i,\cdot}\btheta\right)\}^{1-y_i}\Phi\left((\bX_0)_{i,\cdot}\btheta\right)^{y_i}\right]}
{\ds \mathbb{E}_{\mathcal{N}_3(0_3,n(\bX_1^\text{T}\bX_1)^{-1})}\left[\prod_{i=1}^n 
\{1-\Phi\left((\bX_1)_{i,\cdot}\btheta\right)\}^{1-y_i}\Phi\left((\bX_1)_{i,\cdot}
\btheta\right)^{y_i}\right]}\nonumber
\end{align}
using the shortcut notation that $A_{i,\cdot}$ is the $i$-th line of the matrix $A$.

\section{The basic Monte Carlo solution}
As already shown above, when testing 
for a null hypothesis (or a model) $H_0:\theta\in\Theta_0$ against the alternative hypothesis
(or the alternative model) $H_1:\theta\in\Theta_1$, the Bayes factor is defined by
$$
B_{01}(y)={\displaystyle{\int_{\Theta_0} f(y|\theta_0) \pi_0(\theta_0) \text{d}\theta_0} }\bigg/
{\displaystyle{\int_{\Theta_1} f(y|\theta_1) \pi_1(\theta_1) \text{d}\theta_1} }\,.
$$
We assume in this survey that the prior distributions under both the null and the alternative hypotheses are 
proper, as, typically, they should be. (In the case of common nuisance parameters, a common improper prior
measure can be used on those, see \cite{berger:pericchi:varshavsky:1998,marin:robert:2007}. This obviously
complicates the computational aspect, as some methods like crude Monte Carlo cannot be used at all, while others
are more prone to suffer from infinite variance.)
In that setting, the most elementary approximation to $B_{01}(y)$ consists in using a ratio of two
standard Monte Carlo approximations based on simulations from the corresponding priors. Indeed, for 
$i=0,1$:
$$
\int_{\Theta_i} f(y|\theta) \pi_i(\theta) \text{d}\theta=\mathbb{E}_{\pi_i}\left[f(y|\theta)\right]\,.
$$
If $\theta_{0,1},\ldots,\theta_{0,n_0}$ and $\theta_{1,1},\ldots,\theta_{1,n_1}$
are two independent samples generated from the prior distributions $\pi_0$ and $\pi_1$, respectively, 
then
\begin{equation}
\frac{n_0^{-1}\sum_{j=1}^{n_0}f(y|\theta_{0,j})}{n_1^{-1}\sum_{j=1}^{n_1}f(y|\theta_{1,j})}
\label{eq:bfmc}
\end{equation}
is a strongly consistent estimator of $B_{01}(y)$.

In most cases, sampling from the prior distribution corresponding to either hypothesis is straightforward and
fast. Therefore, the above estimator is extremely easy to derive as a brute-force evaluation of the
Bayes factor. However, if any of the posterior distributions
is quite different from the corresponding prior distribution---and it should be for vague priors---,
the Monte Carlo evaluation of the corresponding evidence is highly inefficient since the sample will be
overwhelmingly producing negligible values of $f(y|\theta_{i,j})$. In addition, if $f^2(y|\theta)$ is not
integrable against $\pi_0$ or $\pi_1$, the resulting estimation has an infinite variance. Since importance
sampling usually requires an equivalent computation effort, with a potentially highy efficiency reward,
crude Monte Carlo approaches of this type are usually disregarded.

\begin{figure}
\centerline{\includegraphics[width=10cm,height=5cm]{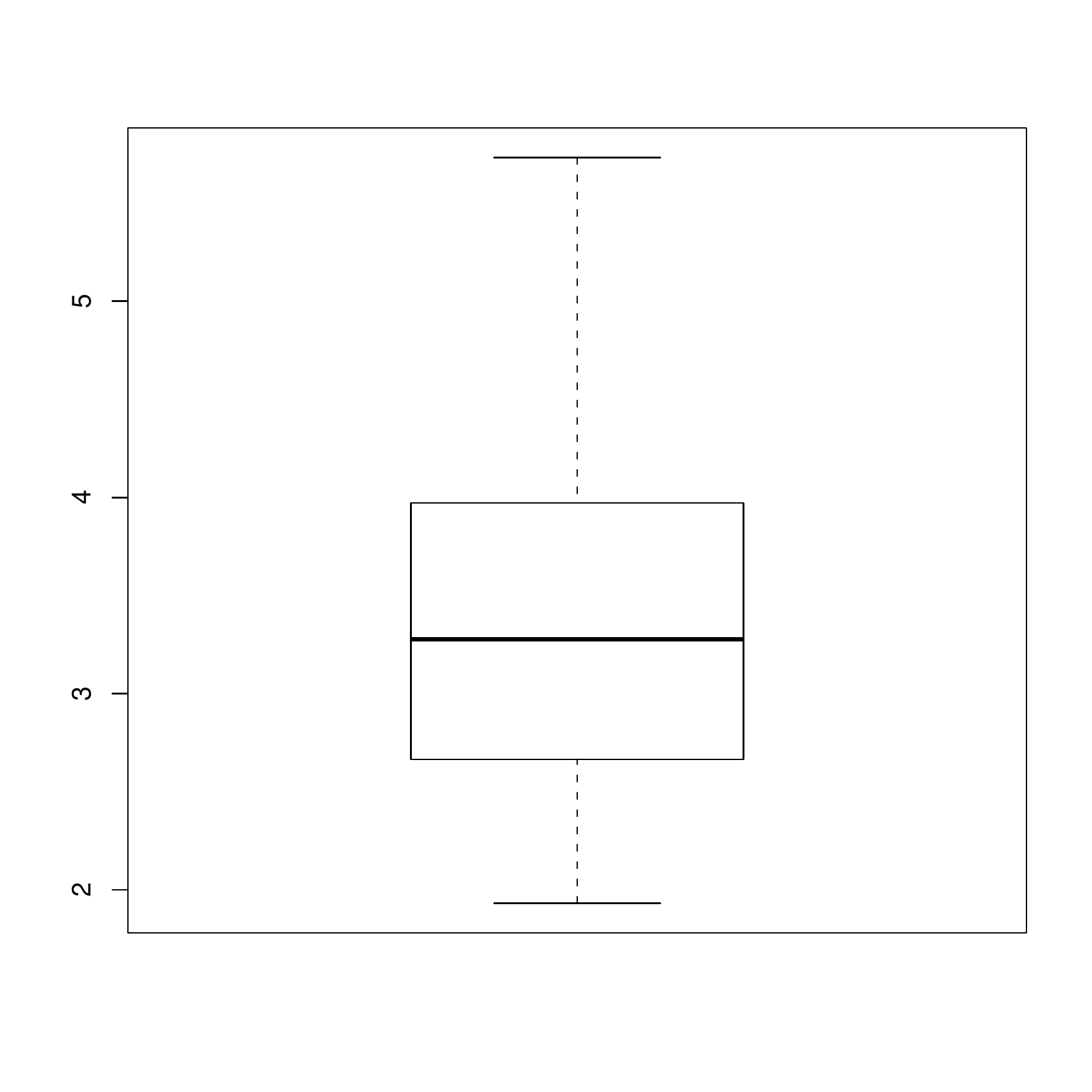}}
\caption{\label{fig:bfmc} Pima Indian dataset: boxplot of 100 Monte Carlo estimates of $B_{01}(y)$
based on simulations from the prior distributions, for $2\times 10^4$ simulations.}
\end{figure}

Figure \ref{fig:bfmc} and Table \ref{tab:res} summarize the results based on $100$ replications 
of Monte Carlo approximations of $B_{01}(y)$,
using equation (\ref{eq:bfmc}) with $n_0=n_1=20,000$ simulations. 
As predicted, the variability of the estimator is very high, when compared with the
other estimates studied in this survey. (Obviously, the method is asymptotically unbiased and, the
functions being square integrable in \eqref{eq:bfprobit}, with a finite variance. A
massive simulation effort would obviously lead to a precise estimate of the Bayes factor.)

\section{Usual importance sampling approximations}
Defining two importance distributions with densities $\varpi_0$ and $\varpi_1$, with 
the same supports as $\pi_0$ and $\pi_1$, respectively, we have:
$$
B_{01}(y)={\mathbb{E}_{\varpi_0}\left[f(y|\theta)\pi_0(\theta)\big/\varpi_0(\theta)\right]}
\bigg/
{\mathbb{E}_{\varpi_1}\left[f(y|\theta)\pi_1(\theta)\big/\varpi_1(\theta)\right]}\,.
$$
Therefore, given two independent samples generated from distributions $\varpi_0$ and $\varpi_1$,
$\theta_{0,1},\ldots,\theta_{0,n_0}$ and $\theta_{1,1},\ldots,\theta_{1,n_1}$, respectively,
the corresponding importance sampling estimate of $B_{01}(y)$ is
\begin{equation}
\dfrac{n_0^{-1} \sum_{j=1}^{n_0} f_0(x|\theta_{0,j}) \pi_0(\theta_{0,j})/\varpi_0(\theta_{0,j})}
{n_1^{-1} \sum_{j=1}^{n_1} f_1(x|\theta_{1,j} \pi_1(\theta_{1,j})/\varpi_1(\theta_{1,j})}\,.
\label{eq:bfis}
\end{equation}
Compared with the standard Monte Carlo approximation above, this approach offers
the advantage of opening the choice of the representation \eqref{eq:bfis} in that 
it is possible to pick importance distributions $\varpi_0$ and $\varpi_1$ that lead
to a significant reduction in the variance of the importance sampling estimate. This implies
choosing importance functions that provide as good as possible approximations to the
corresponing posterior distributions. Maximum likelihood asymptotic distributions or
kernel approximations based on a sample generated from the posterior are natural candidates
in this setting, even though the approximation grows harder as the dimension increases.

\begin{figure}
\centerline{\includegraphics[width=10cm,height=5cm]{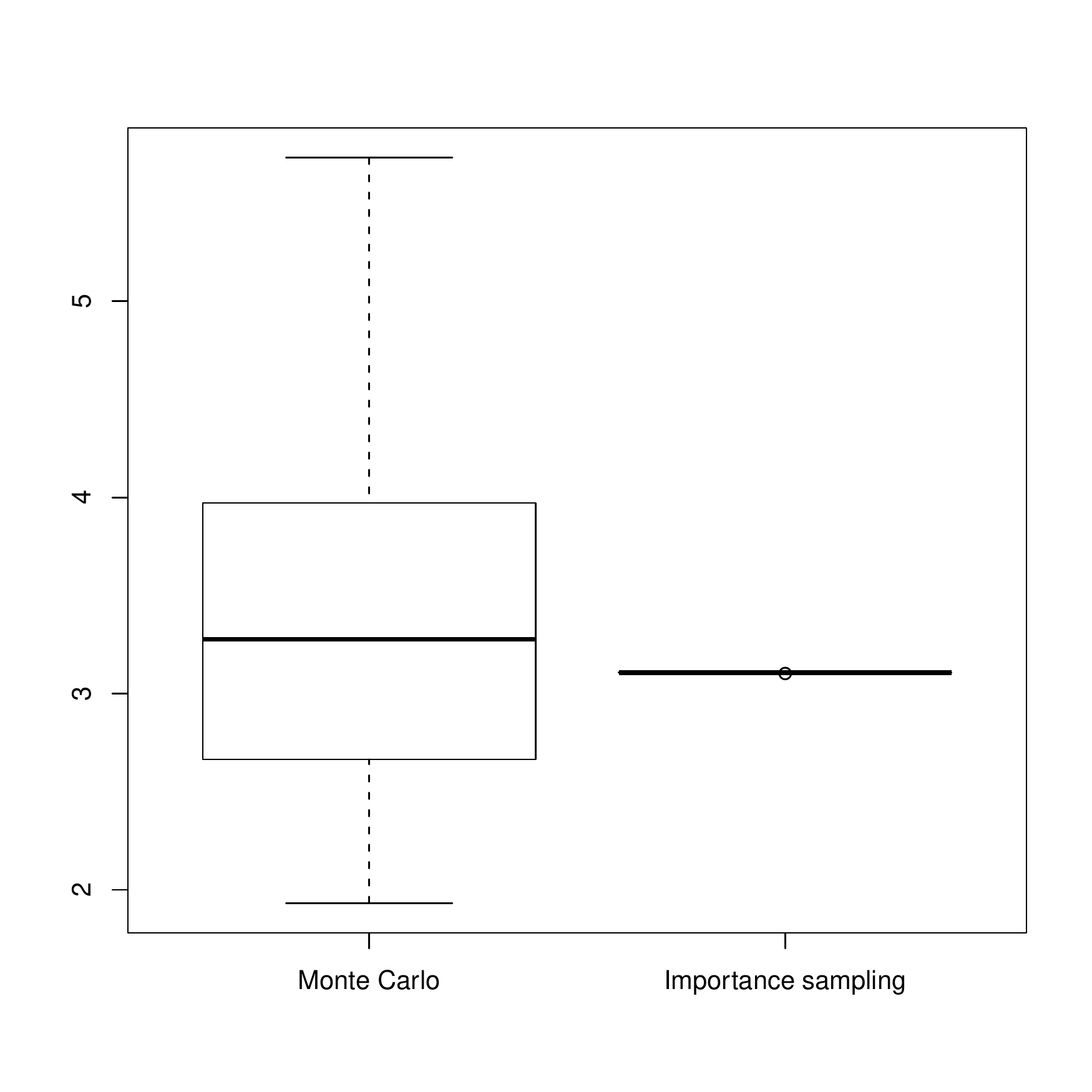}}
\caption{\label{fig:bfmcis} Pima Indian dataset: boxplots of 100 Monte Carlo and 
importance sampling estimates of $B_{01}(y)$,
based on simulations from the prior distributions, for $2\times 10^4$ simulations.}
\end{figure}

For the Pima Indian benchmark, we propose for instance to use as importance distributions,
Gaussian distributions with means equal to the maximum likelihood (ML) estimates
and covariance matrices equal to the estimated covariance matrices of the ML estimates,
both of which are provided by the {\sf R glm()} function.
While, in general, those Gaussian distributions provide crude approximations to the posterior
distributions, the specific case of the probit model will show this is an exceptionally
good approximation to the posterior, since this leads to the best solution among all those
compared here. The results, obtained over $100$ replications
of the methodology with $n_0=n_1=20,000$ are summarized in Figure \ref{fig:bfmcis} and
Table \ref{tab:res}. They are clearly excellent, while requiring the same computing time as
the original simulation from the prior. 

\section{Bridge sampling methodology}

The original version of the bridge sampling approximation to the Bayes factor  
\citep{gelman:meng:1998,chen:shao:ibrahim:2000} relies on the assumption
that the parameters of both models under comparison belong to the same space:
$\Theta_0=\Theta_1$. In that case, for likelihood functions $f_0$ and $f_1$ under respectively models
$\mathfrak{M}_0$ and $\mathfrak{M}_1$, the bridge representation of the Bayes factor is
\begin{equation}
B_{01}(y)={\displaystyle{\int_{\Theta_0} f_0(y|\theta) \pi_0(\theta) \text{d}\theta} }\bigg/
{\displaystyle{\int_{\Theta_1} f_1(y|\theta) \pi_1(\theta) \text{d}\theta}}=
\mathbb{E}_{\pi_1}\left[\left.\frac{f_0(y|\theta) \pi_0(\theta)}{f_1(y|\theta) \pi_1(\theta)}\right|y\right]\,.
\label{eq:bridge1}
\end{equation}
Given a sample from the posterior distribution of $\theta$ under
model $\mathfrak{M}_1$, $\theta_{1,1},\ldots,\theta_{1,N}\sim \pi_1(\theta|y)$,
a first bridge sampling approximation to $B_{01}(y)$ is
$$
N^{-1}\sum_{j=1}^{N}\frac{f_0(y|\theta_{1,j}) \pi_0(\theta_{1,j})}{f_1(y|\theta_{1,j}) \pi_1(\theta_{1,j})}\,.
$$
From a practical perspective, for the above bridge sampling
approximation to be of any use, the constraint on the common parameter space for
both models goes further in that,
not only must both models have the same complexity, but they must also be parameterised on a
common ground, i.e.~in terms of some specific moments of the sampling model,
so that parameters under both models have a common meaning. Otherwise, the resulting
bridge sampling estimator will have very poor convergence properties, possibly with infinite
variance.

Equation (\ref{eq:bridge1}) is nothing but a very special case of the general representation
\citep{torrie:valleau:1977}
$$
B_{01}(y) = {\mathbb{E}_\varphi\left[f_0(y|\theta) \pi_0(\theta) / \varphi(\theta)\right]}\bigg/
{\mathbb{E}_\varphi\left[f_1(y|\theta) \pi_1(\theta) / \varphi(\theta)\right]} \,,
$$
which holds for any density $\varphi$ with a sufficiently large support and which only requires a single
sample $\theta_1,\ldots,\theta_{N}$ generated from $\varphi$ to produce an importance sampling estimate
of the ratio of the marginal likelihoods. Apart from using the {\em same} importance function 
$\varphi$ for both integrals, this method is therefore a special case of importance sampling. 

Another extension of this bridge sampling approach is based on the general representation
\begin{eqnarray*}
B_{01}(y) & = & {\ds \int f_0(y|\theta) \pi_0(\theta) \alpha(\theta) {\pi}_1(\theta|y) \text{d}\theta }\bigg/
{\ds \int f_1(y|\theta) \pi_1(\theta) \alpha(\theta) {\pi}_0(\theta|y) \text{d}\theta } \\  
& \approx & 
\frac{\ds {n_1}^{-1} \sum_{j=1}^{n_1} f_0(y|\theta_{1,j}) \pi_0(\theta_{1,j}) \alpha(\theta_{1,j})}
{\ds {n_0}^{-1} \sum_{j=1}^{n_0} f_1(y|\theta_{0,j}) \pi_1(\theta_{0,j}) \alpha(\theta_{0,j})}
\end{eqnarray*}
where $\theta_{0,1},\ldots,\theta_{0,n_0}$ and $\theta_{1,1},\ldots,\theta_{1,n_1}$
are two independent samples coming from the posterior distributions $\pi_0(\theta|y)$
and $\pi_1(\theta|y)$, respectively. That applies for any positive function $\alpha$ as long as
the upper integral exists. Some choices of $\alpha$ lead to very poor performances of the method in connection 
with the harmonic mean approach (see Section \ref{sec:harmo}), but there
exists a quasi-optimal solution, as provided by \cite{gelman:meng:1998}:
$$
{\alpha^\star(y) \propto \dfrac{1}{n_0{\pi}_0(\theta|y) + n_1  {\pi}_1(\theta|y)}} \,.
$$
This optimum cannot be used {\em per se}, since it requires the normalising constants of
both $\pi_0(\theta|y)$ and $\pi_1(\theta|y)$. As suggested by \cite{gelman:meng:1998},
an approximate version uses iterative versions of $\alpha^\star$, based on iterated
approximations to the Bayes factor. Note that this solution recycles simulations from both posteriors, 
which is quite appropriate since one model is selected via the Bayes factor, 
instead of using an importance weighted sample common to
both approximations. We will see below an alternative representation of the bridge factor 
that bypasses this difficulty (if difficulty there is!).

Those derivations are, however, restricted to the case where both models have the same complexity and thus 
they do not apply to embedded models, when $\Theta_0\subset\Theta_1$ in such a way that 
$\theta_1=(\theta,\psi)$, i.e., when the submodel corresponds to a specific value $\psi_0$ of $\psi$:
$f_0(y|\theta)=f(y|\theta,\psi_0)$.

The extension of the most advanced bridge sampling strategies to such cases 
requires the introduction of a {\em pseudo-posterior density,} $\omega(\psi|\theta,y)$, on the
parameter that does not appear in the embedded model,
in order to reconstitute the equivalence between both parameter spaces. Indeed, 
if we augment $\pi_0(\theta|y)$ with $\omega(\psi|\theta,y)$, we obtain a joint 
distribution with density $\pi_0(\theta|y)\times\omega(\psi|\theta,y)$ on $\Theta_1$.
The Bayes factor can then be expressed as
\begin{equation}\label{eq:psudo}
B_{01} (y) = 
\dfrac{\ds \int_{\Theta_1} f(y|\theta,\psi_0) \pi_0(\theta)\alpha(\theta,\psi)\pi_1(\theta,\psi|y) 
\text{d}\theta\omega(\psi|\theta,y)\,\text{d}\psi}
{\ds \int_{\Theta_1} f(y|\theta,\psi) \pi_1(\theta,\psi) \alpha(\theta,\psi)\pi_0(\theta|y) 
\times\omega(\psi|\theta,y) \text{d}\theta \,\text{d}\psi}\,,   
\end{equation}
for all functions $\alpha(\theta,\psi)$,
because it is clearly independent from the choice of both $\alpha(\theta,\psi)$ and $\omega(\psi|\theta,y)$. 
Obviously, the performances of the approximation
$$
\dfrac{\ds (n_1)^{-1} \sum_{j=1}^{n_1} f(y|\theta_{1,j},\psi_0) 
\pi_0(\theta_{1,j}) \omega(\psi_{1,j}|\theta_{1,j},y)\alpha(\theta_{1,j},\psi_{1,j})}
{\ds (n_0)^{-1} \sum_{j=1}^{n_0} f(y|\theta_{0,j},\psi_{0,j}) \pi_1(\theta_{0,j},
\psi_{0,j})  \alpha(\theta_{0,j},\psi_{0,j})}\,,
$$
where $(\theta_{0,1},\psi_{0,1}),\ldots,(\theta_{0,n_0},\psi_{0,n_0})$ and 
$(\theta_{1,1},\psi_{1,1}),\ldots,(\theta_{1,n_1},\psi_{1,n_1})$
are two independent samples generated from distributions $\pi_0(\theta|y)\times\omega(\psi|\theta,y)$
and $\pi_1(\theta,\psi|y)$, respectively, do depend on this completion by the pseudo-posterior
as well as on the function $\alpha(\theta,\psi)$.
\cite{chen:shao:ibrahim:2000} establish that the asymptotically optimal choice for 
$\omega(\psi|\theta,y)$ is the obvious one, namely
$$
\omega(\psi|\theta,y) = \pi_1(\psi|\theta,y) \,,
$$
which most often is unavailable in closed form (especially when considering that 
the normalising constant of $\omega(\psi|\theta,y)$ is required in \eqref{eq:psudo}). However, in 
latent variable models, approximations of the conditional posteriors often are available, 
as detailed in Section \ref{sec:chibberies}.

While this extension of the basic bridge sampling approximation is paramount for handling embedded models, its implementation suffers
from the dependence on this pseudo-posterior. In addition, this technical device brings the extended bridge methodology
close to the cross-model alternatives of \cite{carlin:chib:1995} and \cite{green:1995}, in that both 
those approaches rely on completing distributions, either locally
\citep{green:1995} or globally \citep{carlin:chib:1995}, to link both models under comparison in a bijective relation. The density
$\omega(\psi|\theta_0,y)$ is then a pseudo-posterior distribution in Chib and Carlin's (1995) sense, and it can be used as
Green's (1995) proposal in the reversible jump MCMC step to move (or not) from model 
$\mathfrak{M}_0$ to model $\mathfrak{M}_1$. While using 
cross-model solutions to compare only two models does seem superfluous, 
given that the randomness in picking the model at each step of the simulation
is not as useful as in the setting of comparing a large number or an infinity of models, the average acceptance 
probability for moving from model $\mathfrak{M}_0$ to model $\mathfrak{M}_1$ is related to the Bayes factor since
$$
\mathbb{E}_{\pi_0\times\omega} \left[\frac{f(y|\theta,\psi)
\pi_1(\theta,\psi)}{f(y|\theta,\psi_0)\pi_0(\theta)\omega(\psi|\theta,y)} \right] = B_{01}(y)
$$
even though the average
$$
\mathbb{E}_{\pi_0\times\omega} \left[ \min\left\{
1,\frac{f(y|\theta,\psi)\pi_1(\theta,\psi)}{f(y|\theta,\psi_0)\pi_0(\theta)\omega(\psi|\theta,y)}\right\}\right]
$$
does not provide a closed form solution.

\begin{figure}
\centerline{\includegraphics[width=5cm,height=5cm]{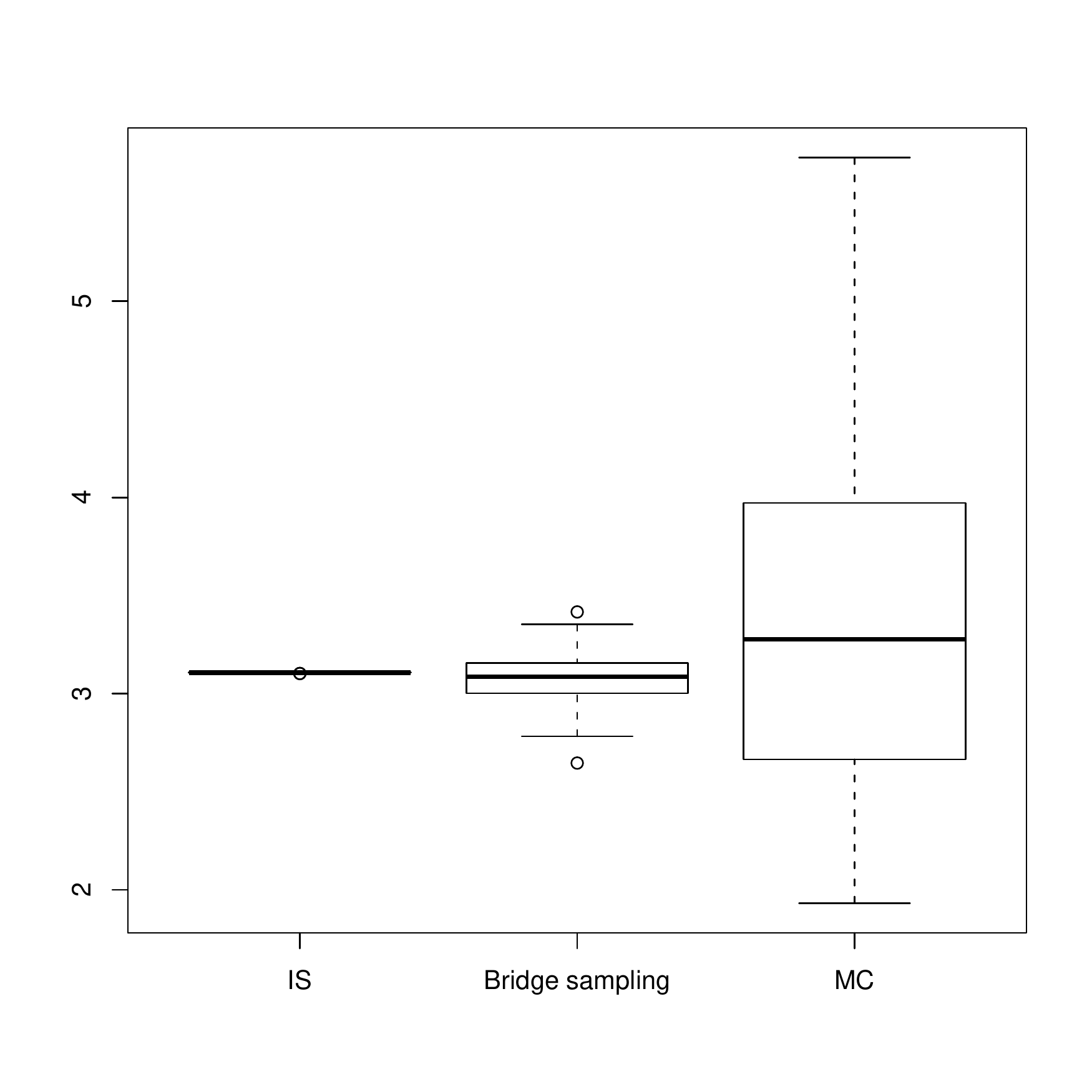}\includegraphics[width=5cm,height=5cm]{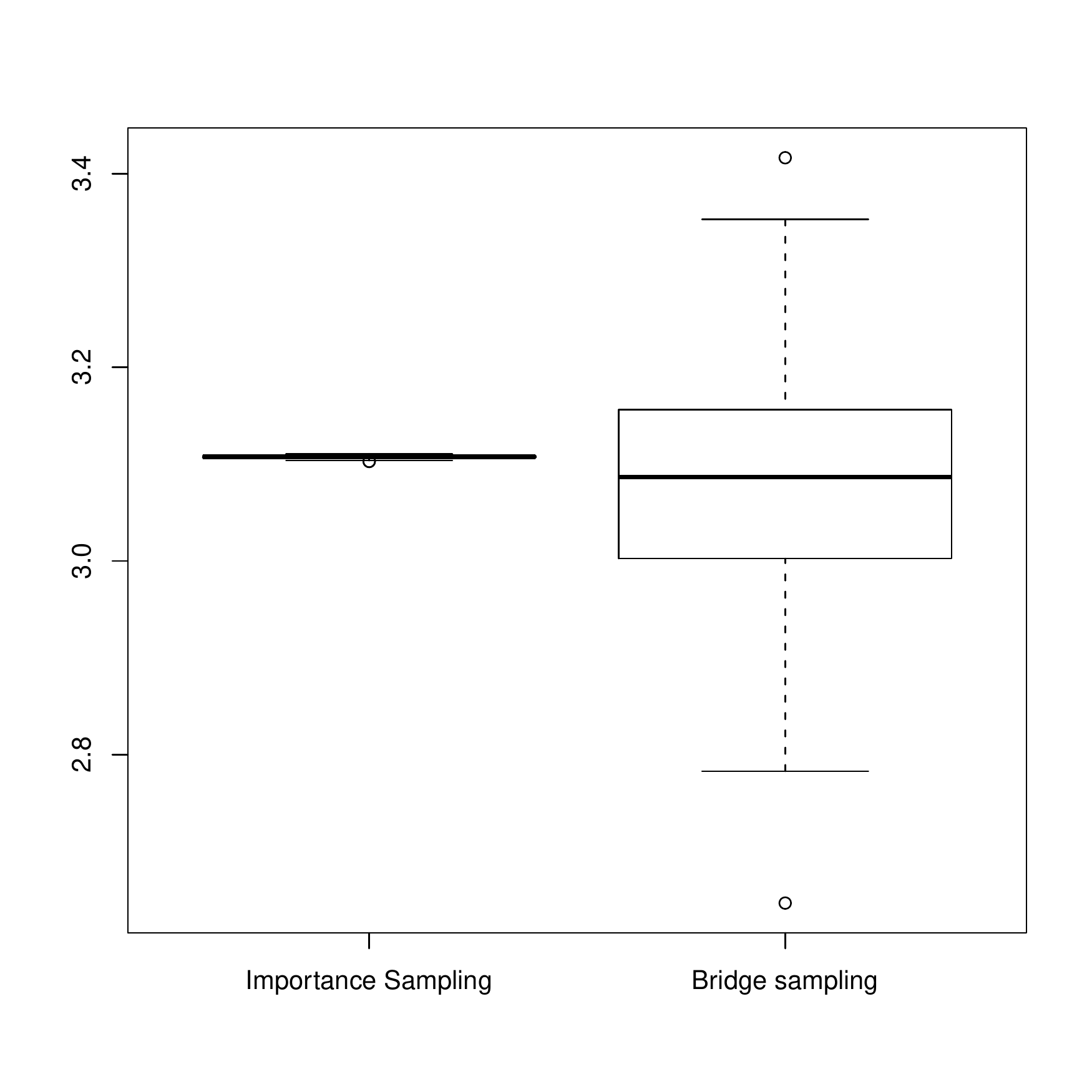}}
\caption{\label{fig:bfbs} Pima Indian dataset: {\em (left)} boxplots of 100 importance sampling, bridge
sampling and Monte Carlo estimates of $B_{01}(y)$, 
based on simulations from the prior distributions, for $2\times 10^4$ simulations;
{\em (right)} same comparison for the importance sampling versus bridge sampling estimates only.}
\end{figure}

For the Pima Indian benchmark, we use as pseudo-posterior density $\omega(\theta_3|\theta_1,\theta_2,y)$,
the conditional Gaussian density deduced from the asymptotic Gaussian distribution on $(\theta_1,\theta_2,\theta_3)$ already
used in the importance sampling solution,
with mean equal to the ML estimate of $(\theta_1,\theta_2,\theta_3)$ and 
with covariance matrix equal to the estimated covariance matrix of the ML estimate. 
The quasi-optimal solution $\alpha^\star$ in the bridge sampling estimate is replaced with the inverse of an average between
the asymptotic Gaussian distribution in model $\mathfrak{M}_1$ and the product of the 
asymptotic Gaussian distribution in model $\mathfrak{M}_0$ times the above $\omega(\theta_3|\theta_1,\theta_2,y)$. This
obviously is a suboptimal choice, but it offers the advantage of providing a non-iterative solution. 
The results, obtained over $100$ replications
of the methodology with $n_0=n_1=20,000$ are summarized in Figure \ref{fig:bfbs} and
Table \ref{tab:res}. The left-hand graph shows that this choice of bridge sampling estimator produces a solution
whose variation is quite close to the (excellent) importance sampling solution, a considerable improvement upon the
initial Monte Carlo estimator. However, the right-hand-side graph shows that the importance sampling solution remains
far superior, especially when accounting for the computing time.
(In this example, running
20,000 iterations of the Gibbs sampler for the models with both two and
three variables takes approximately 32 seconds.)

\section{Harmonic mean approximations}\label{sec:harmo}

While using the generic harmonic mean approximation to the marginal likelihood is often fraught with danger
\citep{neal:1994}, the representation \citep{gelfand:dey:1994} $(k=0,1)$
\begin{equation}\label{eq:harmony}
\mathbb{E}_{\pi_k}\left[\left.\frac{\varphi_k(\theta) }{\pi_k(\theta)f_k(y|\theta)}\right| y \right]
= \int \frac{\varphi_k(\theta) }{\pi_k(\theta)f_k(y|\theta)} \,
\frac{\pi_k(\theta)f_k(y|\theta)}{\ev_k(y)}\,\text{d}\theta
= \frac{1}{\ev_k(y)}
\end{equation}
holds, no matter what the density $\varphi_k(\theta)$ is---provided $\varphi_k(\theta)=0$ when $\pi_k(\theta)f_k(y|\theta)=0$---. 
This representation is remarkable in that it allows for a direct processing of Monte Carlo or MCMC output from the posterior distribution
$\pi_k(\theta|y)$. As with importance sampling approximations, the variability of the corresponding estimator of $B_{01}(y)$ will
be small if the distributions $\varphi_k(\theta)$ ($k=0,1$) are close to the corresponding posterior distributions. However,
as opposed to usual importance sampling constraints, the density
$\varphi_k(\theta)$ must have lighter---rather than fatter---tails than $\pi_k(\theta)f_k(y|\theta)$ 
for the approximation of the marginal $m_k(y)$
$$
1\Bigg/ N^{-1}\,\sum_{j=1}^N \frac{\varphi_k(\theta_{k,j})}{\pi_k(\theta_{k,j}) f_k(y|\theta_{k,j})}
$$
to enjoy finite variance. For instance, using $\varphi_k(\theta)=\pi_k(\theta)$ as in the original harmonic mean approximation
\citep{newton:raftery:1994} will most usually result in an infinite variance estimator, as discussed by \cite{neal:1994}. 
On the opposite, using $\varphi_k$'s with constrained supports derived from a Monte Carlo sample,
like the convex hull of the simulations corresponding to the $10\%$ or to the $25\%$ HPD regions---that again is easily derived
from the simulations---is both completely appropriate and implementable \citep{robert:wraith:2009}.

\begin{figure}
\centerline{\includegraphics[width=5cm,height=5cm]{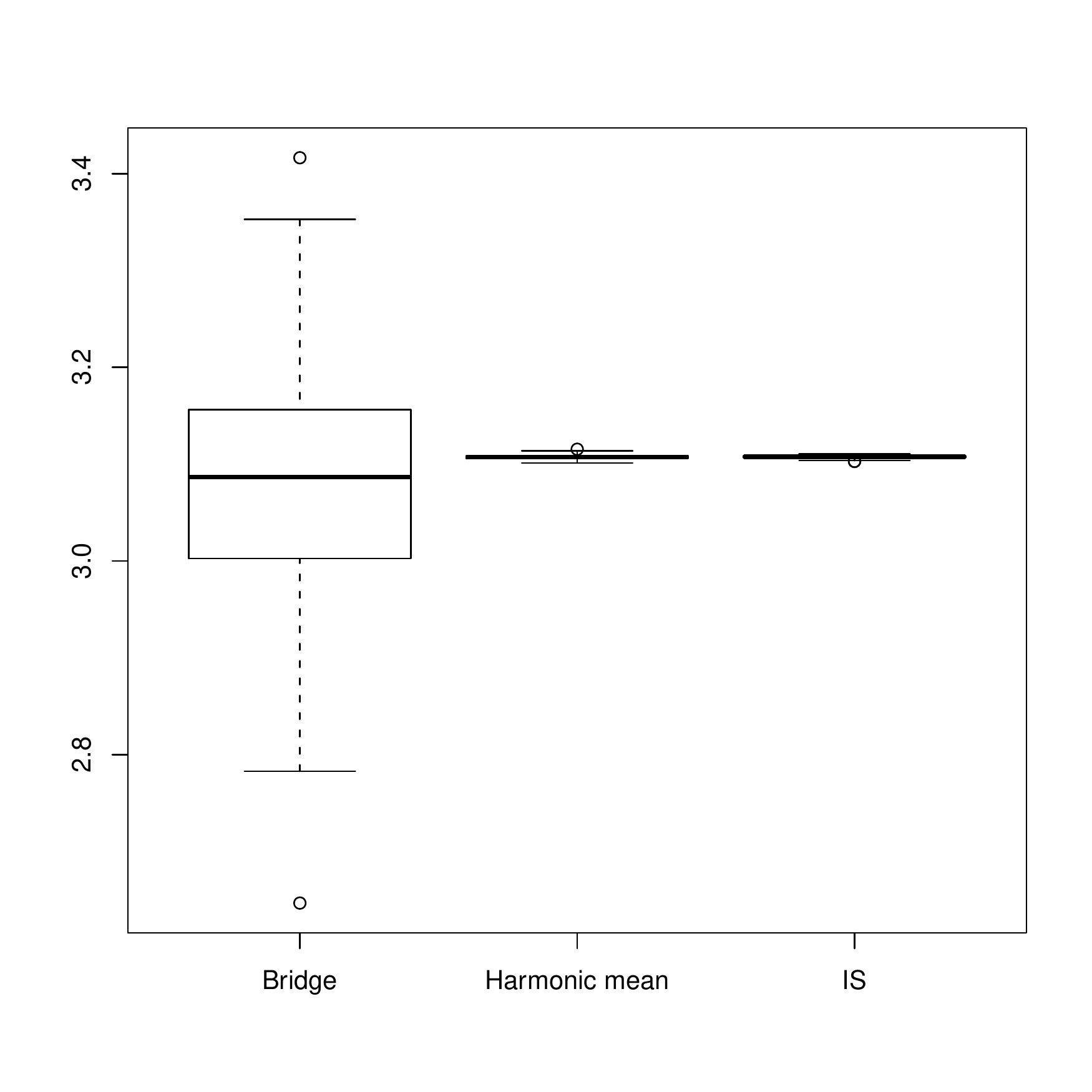}\includegraphics[width=5cm,height=5cm]{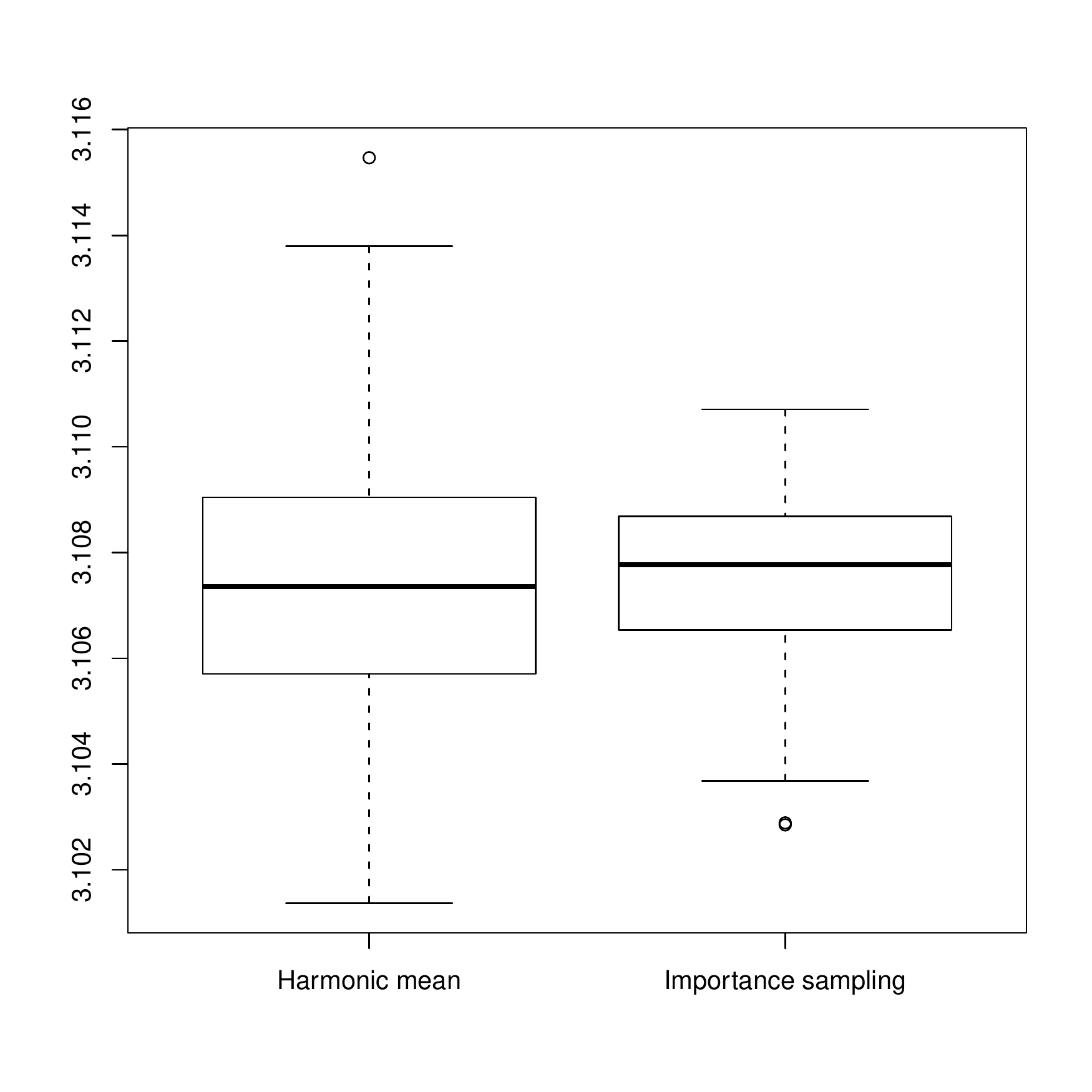}}
\caption{\label{fig:bfhm} Pima Indian dataset: {\em (left)} boxplots of 100 
bridge sampling, harmonic mean and importance sampling estimates of $B_{01}(y)$,
based on simulations from the prior distributions, for $2\times 10^4$ simulations;
{\em (right)} same comparison for the harmonic mean versus importance sampling estimates only.}
\end{figure}

However, for the Pima Indian benchmark, we propose to use instead as our distributions $\varphi_k(\theta)$ the 
very same distributions as those used in the above importance sampling approximations, that is,
Gaussian distributions with means equal to the ML estimates and covariance matrices equal to the estimated covariance
matrices of the ML estimates. The results, obtained over $100$ replications
of the methodology with $N=20,000$ simulations for each approximation of $\ev_k(y)$ ($k=0,1$) are summarized in Figure
\ref{fig:bfhm} and Table \ref{tab:res}. They show a very clear proximity between both importance solutions in this
special case and a corresponding domination of the bridge sampling estimator, even though the importance sampling
estimate is much faster to compute. This remark must be toned down by considering that the computing time due to the
Gibbs sampler should not necessarily be taken into account into the comparison, since samples are generated under both
models.

\section{Exploiting functional equalities}\label{sec:chibberies}

Chib's (1995) method for approximating a marginal (likelihood) is a direct application of
Bayes' theorem: given $y\sim f_k(y|\theta)$ and $\theta\sim\pi_k(\theta)$, we have that
$$
\ev_k = \frac{f_k(y|\theta)\,\pi_k(\theta)}{\pi_k(\theta|y)}\,,
$$
for all $\theta$'s (since both the lhs and the rhs of this equation are constant in $\theta$). Therefore, if an
arbitrary value of $\theta$, say $\theta^*_k$, is selected and if a good approximation to $\pi_k(\theta|y)$
can be constructed, denoted $\hat{\pi}(\theta|y)$, Chib's (\citeyear{chib:1995})
approximation to the evidence is
\begin{equation}\label{eq:chib}
\ev_k = \frac{f_k(y|\theta^*_k)\,\pi_k(\theta^*_k)}{\hat{\pi_k}(\theta^*_k|y)}\,.
\end{equation}
In a general setting, $\hat{\pi}(\theta|y)$ may be the Gaussian approximation based on the MLE, already
used in the importance sampling, bridge sampling and harmonic mean solutions, but this is unlikely to be accurate in a
general framework.
A second solution is to use a nonparametric approximation based on a preliminary MCMC sample, even though the accuracy may
also suffer in large dimensions.  In the special setting of latent variables models (like mixtures of distributions
but also like probit models), Chib's (1995) approximation is particularly attractive as there exists
a natural approximation to $\pi_k(\theta|y)$, based on the Rao--Blackwell
\citep{gelfand:smith:1990} estimate
$$
\hat{\pi_k}(\theta^*_k|y) = \frac{1}{T}\,\sum_{t=1}^T \pi_k(\theta^*_k|y,z_k^{(t)})\,,
$$
where the $z_k^{(t)}$'s are the latent variables simulated by the MCMC sampler.
The estimate $\hat{\pi_k}(\theta^*_k|y)$
is a parametric unbiased approximation of $\pi_k(\theta^*_k|y)$ that converges with rate $\text{O}(\sqrt{T})$.
This Rao--Blackwell approximation obviously requires the full conditional density $\pi_k(\theta^*_k|y,z)$ to be
available in closed form (constant included) but, as already explained, this is the case for the probit model.

\begin{figure}
\centerline{\includegraphics[width=10cm,height=5cm]{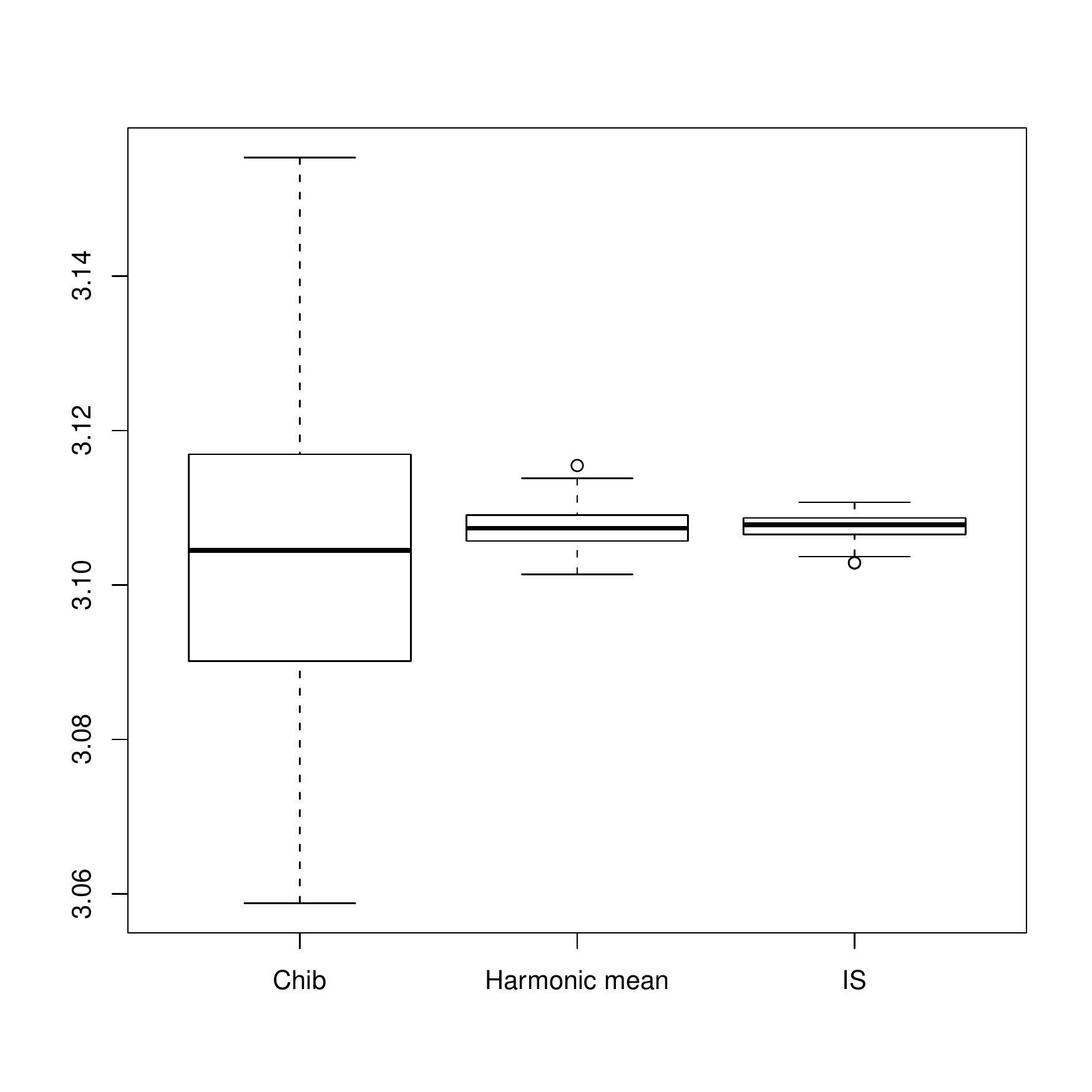}}
\caption{\label{fig:bfchi} Pima Indian dataset: boxplots of 100 Chib's, harmonic mean and importance
estimates of $B_{01}(y)$, based on simulations from the prior distributions, for $2\times 10^4$ simulations.}
\end{figure}

Figure \ref{fig:bfchi} and Table \ref{tab:res} summarize the results obtained for $100$ replications of Chib's approximations
of $B_{01}(y)$ with $T=20,000$ simulations for each approximation of $\ev_k(y)$ ($k=0,1$). While Chib's method is usually
very reliable and dominates importance sampling, the incredibly good approximation provided by the asymptotic Gaussian
distribution implies that, in this particular case, Chib's method is dominated by both the importance sampling and the
harmonic mean estimates.

{\small
\begin{center}
\begin{table}[hbt]
\caption{\label{tab:res} Pima Indian dataset:
Performances of the various approximation methods used in this survey.}
\begin{tabular}{l|l|l|l|l|l|}
                           & Monte  & Importance    & Bridge   & Harmonic & Chib's \\
                           & Carlo  & sampling      & sampling & mean     & approximation \\
\hline Median              & 3.277  & 3.108         & 3.087    & 3.107    & 3.104 \\ 
\hline Standard deviation  & 0.7987 & 0.0017        & 0.1357   & 0.0025   & 0.0195 \\
\hline Duration in seconds & 7      & 7             & 71       & 70       & 64 \\ 
\end{tabular}
\end{table}
\end{center}
}

% Concerning duration in seconds: 20,000 iterations of the Gibbs sampler for models with 2 and
% 3 variables take approximately 32 seconds. \\
% Characteristic of the software used: R 2.9.2 GUI 1.29 Tiger build 32-bit (5464) \\
% Characteristic of the computer used: \\
% System Mac OS X (10.5.8) (Darwin 9.8.0) \\
% Processor 2.4 GHz Intel Core 2 Duo \\
% Ram: 2 Go 1067 MHz DDR3 

\section{Conclusion}

In this short evaluation of the most common estimations to the Bayes factor, we have found that a particular importance
sampling and its symmetric harmonic mean counterpart are both very efficient in the case of the probit model. The bridge
sampling estimate is much less efficient in this example, due to the approximation error resulting from the pseudo-posterior.
In most settings, the bridge sampling is actually doing better than the equivalent importance sampler 
\citep{robert:wraith:2009}, while Chib's method is much more generic than the four alternatives. The recommendation resulting
from the short experiment above is therefore to look for handy approximations to the posterior distribution, whenever 
available, but to fall back on Chib's method as a backup solution providing a reference or better.

%%%%%%%%%%%%%%%%%%%%%%%%%%%%%%%%%%%%%%%%%%%%%%%%
%% BACKMATTER
%%%%%%%%%%%%%%%%%%%%%%%%%%%%%%%%%%%%%%%%%%%%%%%%

\begin{theacknowledgments}
J.-M.~Marin and C.P.~Robert are supported by the 2009--2012 grant ANR-09-BLAN-0218 ``Big'MC".
\end{theacknowledgments}

%%%%%%%%%%%%%%%%%%%%%%%%%%%%%%%%%%%%%%%%%%%
%% Just a reminder that you may have to run bibtex
%% All of it up to \end{document} can be removed
%% if you don't like the warning.
%%%%%%%%%%%%%%%%%%%%%%%%%%%%%%%%%%%%%%%%%%%
\IfFileExists{\jobname.bbl}{}
 {\typeout{}
  \typeout{******************************************}
  \typeout{** Please run "bibtex \jobname" to obtain}
  \typeout{** the bibliography and then re-run LaTeX}
  \typeout{** twice to fix the references!}
  \typeout{******************************************}
  \typeout{}
 }

\end{document}